# Stabilization of NbTe$_3$, VTe$_3$, and TiTe$_3$ *via* Nanotube Encapsulation

Scott Stonemeyer[1,2,3,4†], Jeffrey D. Cain[1,3,4†], Sehoon Oh[1,4†], Amin Azizi[1,3], Malik Elasha[1], Markus Thiel[1], Chengyu Song[5], Peter Ercius[5], Marvin L. Cohen[1,4], and Alex Zettl[1,3,4]*

[1]*Department of Physics, University of California at Berkeley, Berkeley, CA 94720, USA*

[2]*Department of Chemistry, University of California at Berkeley, Berkeley, CA 94720, USA*

[3]*Kavli Energy NanoSciences Institute at the University of California at Berkeley, Berkeley, CA 94720, USA*

[4]*Materials Sciences Division, Lawrence Berkeley National Laboratory, Berkeley, CA 94720, USA*

[5]*National Center for Electron Microscopy, The Molecular Foundry, One Cyclotron Road, Berkeley, CA 94720 USA*

*†These authors contributed equally*

*Correspondence to: azettl@berkeley.edu

**ABSTRACT**

The structure of MX$_3$ transition metal trichalcogenides (TMTs, with M a transition metal and X a chalcogen) is typified by one-dimensional (1D) chains weakly bound together *via* van der Waals interactions. This structural motif is common across a range of M and X atoms (*e.g.* NbSe$_3$, HfTe$_3$, TaS$_3$), but not all M and X combinations are stable. We report here that three new members of the MX$_3$ family which are not stable in bulk, specifically NbTe$_3$, VTe$_3$, and TiTe$_3$, can be synthesized in the few- to single-chain limit *via* nano-confined growth within the stabilizing cavity of multi-walled carbon nanotubes. Transmission electron microscopy (TEM) and atomic-resolution scanning transmission electron microscopy (STEM) reveal the chain-like nature and the detailed atomic structure. The synthesized materials exhibit behavior unique to few-chain quasi-1D structures, such as multi-chain spiraling and a trigonal anti-prismatic rocking distortion in the single-chain limit. Density functional theory (DFT) calculations provide insight into the crystal structure and stability of the materials, as well as their electronic structure.

**INTRODUCTION**

One of the primary objectives of nanoscience is the precise control over the processing-structure-property relationship intrinsic to traditional materials science. Recently, the concept of dimensionality, and the engineering of materials' structure and properties *via* changes in dimensionality, has emerged as an additional degree of freedom within this paradigm. To this end, the isolation of single atomic-planes (*e.g.* graphene[1] and the transition metal dichalcogenides[2]) and chains (*e.g.* transition metal trichalcogenides[3], TMTs) from quasi-low-dimensional materials has been extremely fruitful. There is an ever-growing list of materials being isolated down to their fundamental van der Waals (vdW) building blocks, and existing low-dimensional materials (*e.g.* nanotubes, graphene) have been excellent templates for the synthesis and stabilization of new low-dimensional materials and structures.[4–6]

Carbon nanotubes (CNTs) and boron nitride nanotubes (BNNTs) have been used as nano reaction vessels for the synthesis of a variety of materials, including elemental metals,[7] halides,[8,9] and chalcogenides.[10] The CNT sheath enables the study of 1D nanostructures that are not air-stable, by protecting them from oxidation. In some cases, the nanoconfined growth can induce the formation of crystal structures and morphologies unrealized in bulk counterparts, such as 1D HgTe,[11,12] helices in 1D cobalt iodide,[13] and twisting in single-chain monochalcogenides.[10] Encapsulation of materials inside small-dimeter nanotubes has become a unique route towards new quasi-1D nanostructures.

The archetypal quasi-1D materials family is the TMTs, commonly referred to as $MX_3$ compounds with M a transition metal and X a chalcogen. Typically for these materials 1D $MX_3$ chains are weakly coupled by interchain vdW interactions to form coherent, but highly anisotropic, three-dimensional crystals. The bulk synthesis and properties of the TMTs has been well explored and they are canonical examples of superconductors and charge density wave materials.[14,15] Recently, TMTs have also been isolated in the few- and single-chain limit, where they exhibit unique spiral chain structure and torsional instabilities, induced by the added dimensional constraint.[3,16] Based on chemical trends, one might expect the quasi-1D crystal structure seen in the TMTs to be present in all $MX_3$ compounds with M=Ti, Zr, Hf, Nb, or Ta, and X=S, Se, or Te. While this holds true for many of the possible combinations (*e.g.* $NbSe_3$, $TaS_3$, $HfTe_3$), specimens of many telluride-based TMTs have not been previously synthesized, and are likely not stable in bulk.[17]

Here, we demonstrate the successful synthesis of three previously unreported $MX_3$ TMT compounds: $NbTe_3$, $VTe_3$, and $TiTe_3$. This is accomplished through nano-confined growth within the cavity of multi-walled carbon nanotubes (MWCNTs). Depending upon the inner diameter of the encapsulating MWCNTs, specimens ranging from many chains, to few chains (2-3), and even single chain, can be isolated and studied. The MWCNT sheath stabilizes the chainlike morphology, enabling synthesis and characterization with transmission electron microscopy (TEM) and aberration-corrected transmission electron microscopy (STEM). It is found that few-chain specimens of the new TMTs can exhibit a coordinated interchain spiraling, while the single-chain limit exhibits a trigonal anti-prismatic (TAP) rocking, behaviors seen previously in the few chain limit of $NbSe_3$ and $HfTe_3$.[3,16] First principles calculations give insight into the integral role that the encapsulating CNT plays in stabilization and provide information regarding the electronic structure of the new materials.

The stabilized chains of $NbTe_3$, $VTe_3$, and $TiTe_3$ are synthesized within CNTs using a procedure similar to that outlined previously for $NbSe_3$ and $HfTe_3$.[3,16] Stoichiometric quantities of powdered transition metal along with Te shot (~450 mg total), together with 1-2 mg of end-opened MWCNTs (CheapTubes, 90% SW-DW CNTs) and ~5 mg/cm$^3$ (ampoule volume) of $I_2$ are sealed under vacuum in a quartz ampoule and heated in a uniform temperature furnace at 520 – 625 ºC for several days, then cooled to room temperature over 3-7 days.

Figure 1 shows high-resolution TEM images of representative samples from all three stabilized species encapsulated within MWCNTs, in which their 1D, chain-like nature is evident. In Figure 1a, ~14 $TiTe_3$ chains are encapsulated within a 3.47 nm-wide (inner diameter) MWCNT (number of enclosed chains estimated based on the carbon nanotube diameter and a close-packing configuration of the chains). Figure 1b exhibits triple-spiraling chains of $VTe_3$ within a 2.46 nm MWCNT, while Figure 1c shows straight triple chains of $NbTe_3$ within a 2.57 nm-wide MWCNT. Figure 1d shows a double chain example of $TiTe_3$ within a 2.04 nm MWCNT and Figure 1e highlights the single-chain limit of $NbTe_3$ encapsulated within a 0.99 nm wide MWCNT. These results demonstrate that a confined growth environment allows for the stabilization and characterization of many-, few-, and single-chain limits of $NbTe_3$, $VTe_3$, and $TiTe_3$.

As seen in the TEM images in Figure 1, $NbTe_3$, $VTe_3$, and $TiTe_3$ adopt very similar behaviors when encapsulated, reminiscent of behaviors reported previously for $NbSe_3$ and $HfTe_3$. Spiraling in the triple-chain limit is evident as shown in Figure 1b and Figure S1. Chemical analyses of the new materials by means of energy-dispersive X-ray spectroscopy (EDS) are shown in Figure 2. Figures 2a, 2b, and 2c show annular dark-field scanning transmission electron microscope (ADF-STEM) images of few-chain specimen of $NbTe_3$, $VTe_3$, and $TiTe_3$, respectively, as well as corresponding EDS line scans. EDS mapping is used to distribute the dose, and the maps are summed vertically to increase the signal to noise. The EDS line scans in Figures 2a, 2b, and 2c confirm distribution of the transition metal, Nb, V, or Ti respectively, and Te atoms across the few-chain crystal within the carbon nanotube. The EDS spectra collected from individual few-chain $NbTe_3$, $VTe_3$, and $TiTe_3$ are presented in Figure S2. The spectra clearly show peaks of Nb and Te for $NbTe_3$, V and Te for $VTe_3$, and Ti and Te for $TiTe_3$.

The new TMT compositions are stable down to the single-chain limit. We examine the single-chain structure of these previously unreported chemistries using atomic-resolution annular dark field (ADF-) STEM imaging and STEM image simulation, highlighted in Figure 3. The STEM simulations are generated from the structures obtained by DFT calculations, as discussed in later sections. The left-hand section of the simulation shows the raw multislice output, and the right-hand section of the simulation incorporates noise from Poisson counting statistics according to the experimental dose.[18] The structure and contrast of atomic species observed in the simulations matches the experimental images for each of the single TMT chains, and highlights the ability of the high angular (HA)-ADF-STEM to distinguish the different atomic species in the single chains. In Figure 3a-c, single chains of $NbTe_3$, $VTe_3$, and $TiTe_3$, are seen encapsulated within a 1.11 nm, 1.02 nm, and 1.05 nm wide double-wall CNT, respectively. Figure 3d shows an atomic model of the single-chain TMT encapsulation process in a single-walled CNT, where the CNT clearly visible in the model may not be readily visible in the STEM images because of the contrast difference of the carbon atoms. The most striking aspect of all three compounds' structure in Figures 3a, 3b, and 3c, is what appears to be the intrachain rocking of the Te ligands along the chain axis. Such a rocking was previously suggested by theory for few- and single-chain $HfTe_3$, but not resolved experimentally due to the spatial resolution necessary to observe such a small structural variation.[16] The enhanced experimental resolution in the present study allows for direct confirmation of such a rocking distortion for $NbTe_3$, $VTe_3$, and TiTe$_3$. The intrachain rocking of the Te ligands distorts the trigonal prismatic (TP) chains into a trigonal antiprismatic (TAP) chain structure. This is relatively unsurprising for the single chain of encapsulated TiTe$_3$, as titanium and hafnium are in the same group on the periodic table, so chemical trends imply this behavior. However, the TAP distortion seen in NbTe$_3$ and VTe$_3$ is surprising, as a previously studied encapsulated TMT in the same group, NbSe$_3$, did not show the TAP distortion, but rather a charge-induced torsional wave (CTW). Therefore, the TAP distortion seen in all three of these isolated structures could be closely related to their stability (or lack thereof) in bulk crystals, or indicate this behavior is innate to the single-chain limit of transition metal tritellurides, which is studied in following sections.

We further investigate the structural makeup and the related TAP distortions of single-chain TiTe$_3$, VTe$_3$, and NbTe$_3$ *via* first-principles calculations based on density functional theory (DFT). As no bulk crystalline data are available for these TMT structures, we construct candidate structures for the chains with various symmetries including the TP and TAP phases (Figure S3). From the constructed candidate structures, the atomic structures are optimized by minimizing the total energy. In the optimization, the atomic positions are relaxed with the fixed value of the distance between the nearest transition metal atoms, $b_0^{MTe3}$ (M=Ti, V, and Nb), extracted from the STEM images shown in Figure 3.

We first consider the atomic and electronic structures of the single chains of TiTe$_3$, VTe$_3$, and NbTe$_3$ isolated in vacuum. The obtained atomic and electronic structures of the TP and TAP single chains isolated in vacuum are shown in Figures S4 and S5, respectively. All three of the TP single chains are metallic with Te bands crossing the Fermi level (Figure S4). For the TAP chains, the short-wavelength rocking distortions of Te ligands from a TP to a TAP unit cell, which was initially observed in the ADF-STEM images highlighted in Figure 3, lead to the split of the Te bands near the chemical potential (Figure S5). As a result, semiconducting band gaps of 0.690 and 0.486 eV in single chains of $TiTe_3$ (Figures S5c and S5d) and $NbTe_3$ (Figures S5k and S5l) are created, respectively, while $VTe_3$ single chain remains metallic (Figures S5g and S5h). Note that the TAP structure of the single-chain $TiTe_3$ in vacuum has 0.652 eV/formula unit (f.u.) lower energy than the TP structure in vacuum, while the TAP structures of the single chains of $VTe_3$ and $NbTe_3$ isolated in vacuum have 0.015 and 0.142 eV/f.u. higher energies than the TP structures in vacuum, respectively.

Next, we investigate the atomic and electronic structures of the single-chain TMT encapsulated inside a (14,0) CNT (indices chosen for convenience). We construct the candidate structures using the atomic positions of the single chains isolated in vacuum and those of the empty CNT. From the candidate structures, the atomic positions of the chains are relaxed by minimizing the total energy, whereas the atomic positions of the CNT are fixed. The obtained atomic and electronic structures of the TP and TAP single chains of TMT species encapsulated in the CNT are shown in Figure S6 and Figure 4, respectively. For the TP chains, all three of the $MX_3$ single chains remain metallic, and the encapsulation does not alter the electronic structure of the chains significantly except for the Fermi level shift due to the charge transfer between the chains and CNT as shown in Figures S4 and S6. Figure 4 shows the obtained atomic structures, the electronic band structure, and the projected density of states (PDOS) for the TAP single-chain $MX_3$ encapsulated in the CNT. For the TAP single chains of the $MX_3$, the encapsulation does not alter the electronic structure significantly, except for the chemical potential shift and the gap opening in the single-chain $VTe_3$ (Figures S5 and 4). The encapsulated $TiTe_3$ and $NbTe_3$ chains remain semiconducting with bandgaps of 0.680, and 0.339 eV, respectively, for the electron transfer within the chains. For $VTe_3$, the encapsulation-driven band repulsion at the Fermi level around the zone boundary, $Z_{VTe3}$, opens a bandgap of 0.062 eV as shown in Figures 4g and 4h, leading to a transition from metal to semiconductor. Note that the TAP structures of single-chain TiTe$_3$ and VTe$_3$ encapsulated in the CNT have 0.648 and 0.016 eV/formula unit (f.u.) lower energy than the TP structure encapsulated in the CNT, respectively, while the TAP structures of single-chains NbTe$_3$ encapsulated in the CNT has 0.162 eV/f.u. higher energies than the TP structures encapsulated in the CNT. We calculate the binding energy, $E_b$, of the single-chain MX$_3$, which is defined as $E_b = E_{MX3}^{chain} + E_{CNT} - E_{MX3/CNT}$, where $E_{MX3}^{chain}$ is the total energy of the isolated single-chain MX$_3$, $E_{CNT}$ is the total energy of an empty CNT isolated in vacuum, and $E_{MX3/CNT}$ is the total energy of the joint system of TP or TAP single-chain MX$_3$ encapsulated inside the CNT. The calculated binding energies of the TAP (TP) single chains of TiTe$_3$, VTe$_3$, and NbTe$_3$ are 1.378 (1.381), 1.165 (1.134), and 1.384 (1.404) eV/f.u., respectively.

Finally, we investigate the stability of the TMT families including these new materials by calculating the Gibbs free energies of formation, $\delta G$, of the MX$_3$, which are defined as $\delta G = \epsilon_{MX3} - n_M \epsilon_M - n_X \epsilon_X$, where $\epsilon_{MX3}$, $\epsilon_M$ and $\epsilon_X$ are the total energies per atom of the bulk MX$_3$, the bulk transition metal (M = Ti, Zr, Hf, V, Nb, and Ta), and the bulk chalcogen (X = S, Se, and Te), respectively, and $n_M$ and $n_X$ are the mole fractions of the M and X atoms, respectively, as shown in Table 1. The calculated $\delta G$ increases in the order of sulfide, selenide, and telluride compounds, possibly due to the size effect of the chalcogen atoms. The calculated $\delta G$ of VTe$_3$ and NbTe$_3$ are -0.156 and -0.312 eV/atom, respectively, notably higher than those of the experimentally observed telluride materials, ZrTe$_3$ and HfTe$_3$, which are -0.653 and -0.522 eV/atom, respectively, while that of TiTe$_3$ is -0.506 eV/atom, comparable with those of ZrTe$_3$ and HfTe$_3$. We compare the stability of an MX$_3$ configuration with those of an MX$_2$ configuration by calculating $\Delta E$, which is defined as $\Delta E = E_{MX3}^{bulk} - E_{MX2}^{bulk} - E_X^{bulk}$, where $E_{MX_3}^{bulk}$, $E_{MX_2}^{bulk}$, and $E_X^{bulk}$ are the total energies of the bulk MX$_3$, MX$_2$ and X, respectively. The calculated $\Delta E$ of TiTe$_3$, VTe$_3$, and NbTe$_3$ are 0.193, 0.292, and 0.110 eV/f.u., respectively, meaning the MX$_3$ configurations are less stable than the MX$_2$ configurations, while those of TaS$_3$, NbSe$_3$, and ZrTe$_3$ are -0.155, -0.063, and -0.011 eV/f.u., respectively, meaning the MX$_3$ configurations have higher or comparable stability compared to the MX$_2$ configurations. Based on the calculated $\delta G$ and $\Delta E$, VTe$_3$ has the lowest $\delta G$ and highest $\Delta E$, implying it is the hardest to achieve in bulk form. However, NbTe$_3$ and TiTe$_3$ are also difficult to realize in their bulk form due to low $\delta G$ and $\Delta E$, respectively. This is consistent with there being no reports of successful bulk synthesis of these materials.

In summary, we have demonstrated the few-chain limit synthesis of three new members of the quasi-1D TMT family, *via* encapsulation within the hollow cavity of MWCNTs. The nanotube sheath stabilizes and protects the materials, allowing access to hitherto unseen compositions and crystal structures, specifically NbTe$_3$, VTe$_3$, and TiTe$_3$. We investigated the structures of the new materials with atomic-resolution electron microscopy, revealing helical twisting of chains in multi-chain specimens and trigonal antiprismatic rocking distortion in single chain specimen. DFT calculations illuminate the electronic properties of single chains of the materials encapsulated in MWCNTs and quantify the stability of these materials in their bulk form and in the single-chain limit. Our study lays further groundwork for the study of confinement-stabilized non-equilibrium materials and associated emergent physical phenomena.

**ACKNOWLEDGMENTS**

**Funding:** This work was primarily funded by the U.S. Department of Energy, Office of Science, Office of Basic Energy Sciences, Materials Sciences and Engineering Division, under Contract No. DE-AC02-05-CH11231 within the sp2-Bonded Materials Program (KC2207) which provided for synthesis of the compounds, TEM and STEM structural characterization, and theoretical modeling. The elemental mapping work was funded by the U.S. Department of Energy, Office of Science, Office of Basic Energy Sciences, Materials Sciences and Engineering Division, under Contract No. DE-AC02-05-CH11231 within the van der Waals Heterostructures Program (KCWF16). Work at the Molecular Foundry (TEAM 0.5 characterization) was supported by the Office of Science, Office of Basic Energy Sciences, of the U.S. Department of Energy under Contract No. DE-AC02-05- CH11231. Support was also provided by the National Science Foundation under Grants No. DMR-1807233, which provided for preparation of opened nanotubes, and No. DMR-1926004, which provided for theoretical calculations of the electronic and structural properties. Computational resources were provided by the DOE at Lawrence Berkeley National Laboratory's NERSC facility and the NSF through XSEDE resources at NICS.

**Author Contributions:** S.S., J.C., and A.Z. conceived the idea; S.S., M.E., and M.T. synthesized the materials; S.S., J.C., A.A., C.S., and P.E. conducted S/TEM studies; S.O. performed DFT calculations; A.Z. and M.L.C. supervised the project; and all authors contributed to the discussion of the results and writing of the manuscript.

**Competing Interests:** Authors have no competing interests.

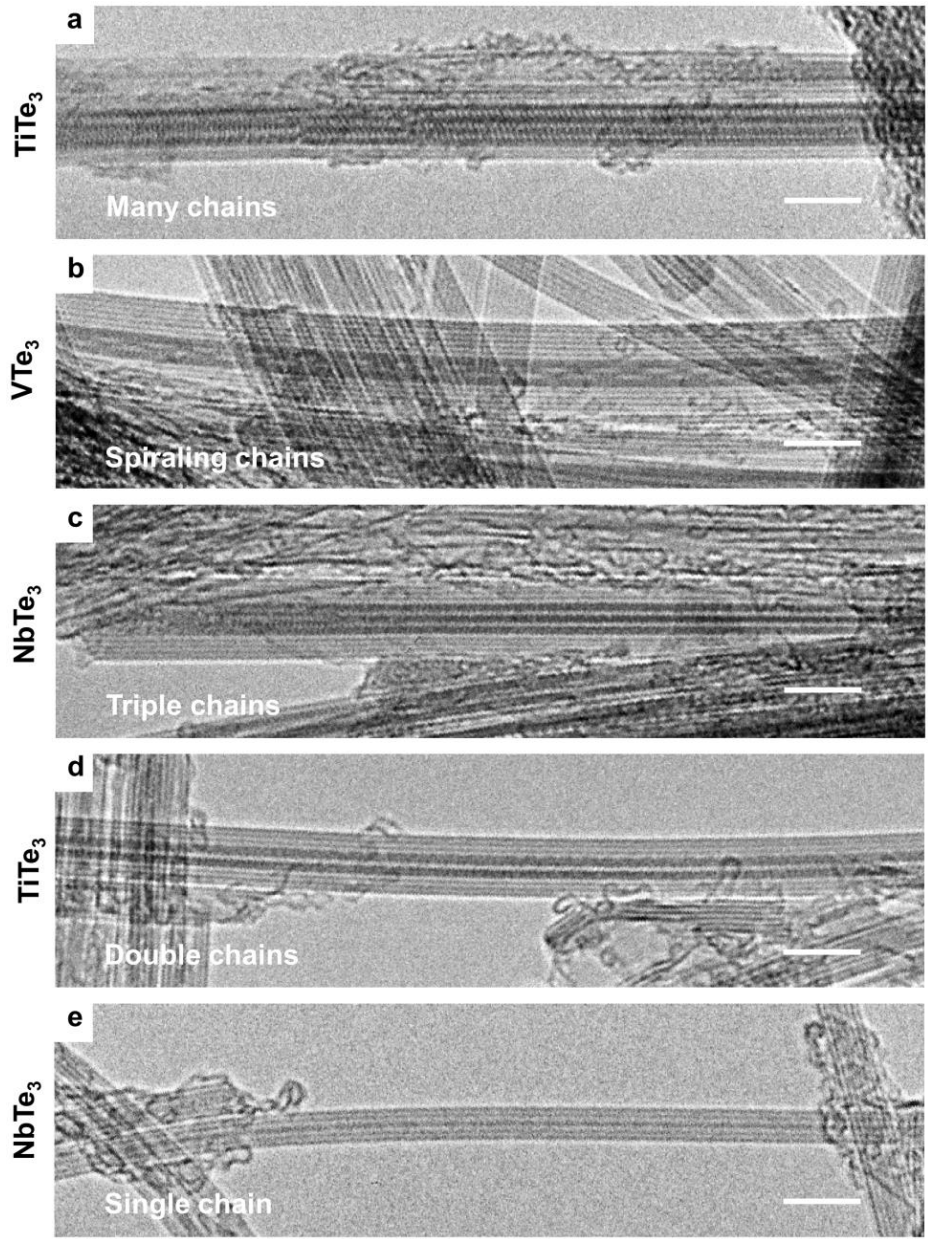

**Figure 1. Representative samples of many- to single-chain limit of TiTe$_3$, VTe$_3$, and NbTe$_3$.** Transmission electron microscopy images of (a) many chains of TiTe$_3$, (b) triple spiraling chains of VTe$_3$, (c) triple chains of NbTe$_3$, (d) double chains of TiTe$_3$, and (e) the single-chain limit of NbTe$_3$. All images are underfocused, where atoms appear dark. Scale bars measure 5 nm.

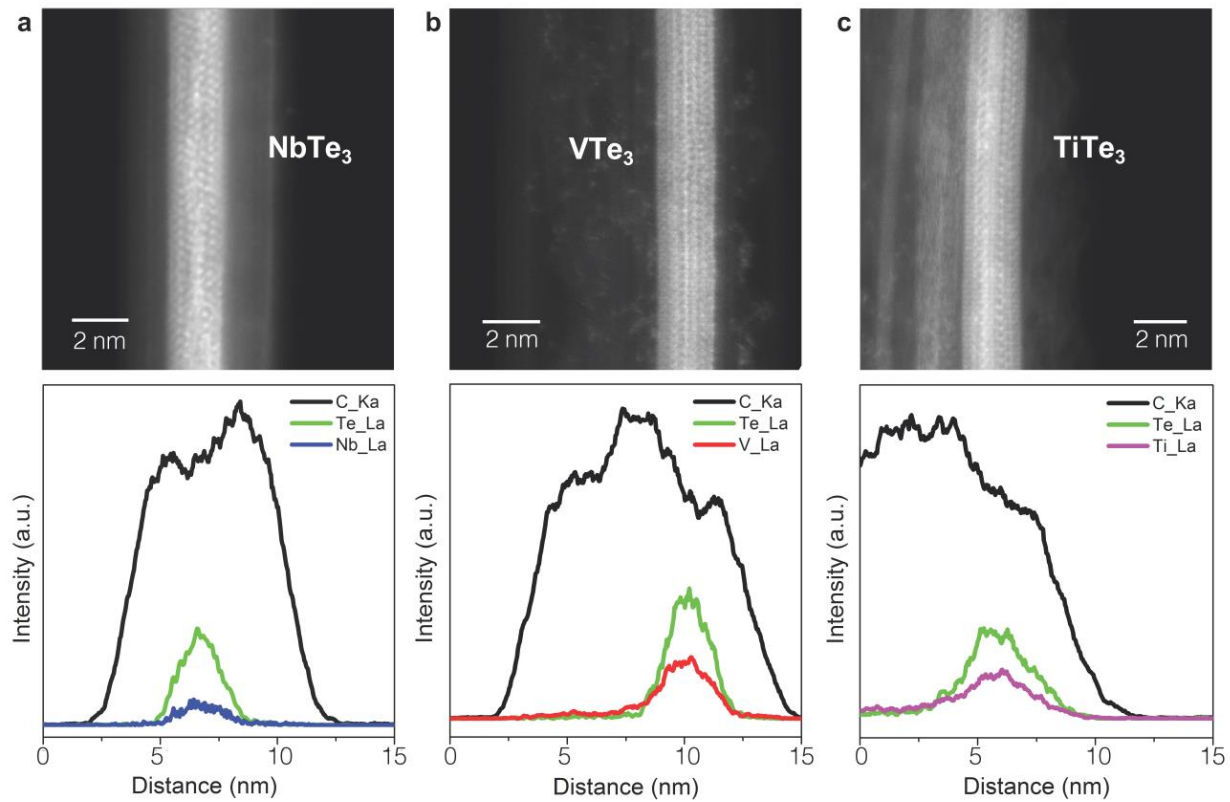

**Figure 2. Elemental analysis of encapsulated few-chain NbTe$_3$, VTe$_3$, and TiTe$_3$.** ADF-STEM images of few-chain specimen and corresponding EDS line scans (summed vertically) for (a) NbTe$_3$, (b) VTe$_3$, and (c) TiTe$_3$. EDS line scans show distribution of the transition metal and tellurium carbon within the carbon nanotube.

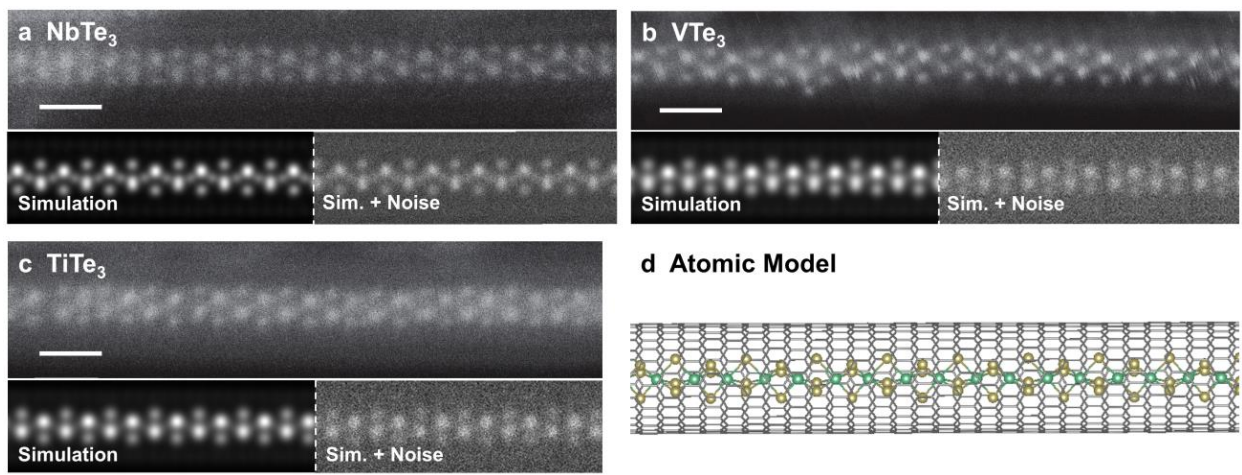

**Figure 3. Single-chain STEM imaging, STEM simulation, and atomic structure of NbTe₃, VTe₃, and TiTe₃.** STEM images of single chains of encapsulated (a) NbTe₃, (b) VTe₃, (c) TiTe₃. Below each STEM image is the corresponding STEM image simulations (left), with appropriate noise from microscope conditions added (right), from the structures obtained by DFT calculations (see Figure 4). Scale bars measure 1 nm. (d) Atomic structure and schematic of a single chain of transition metal tritelluride encapsulated within a carbon nanotube, where the gold and green atoms represent Tellurium and the corresponding transition metal (Nb, V, or Ti), respectively, and the gray lattice represents the encapsulating CNT.

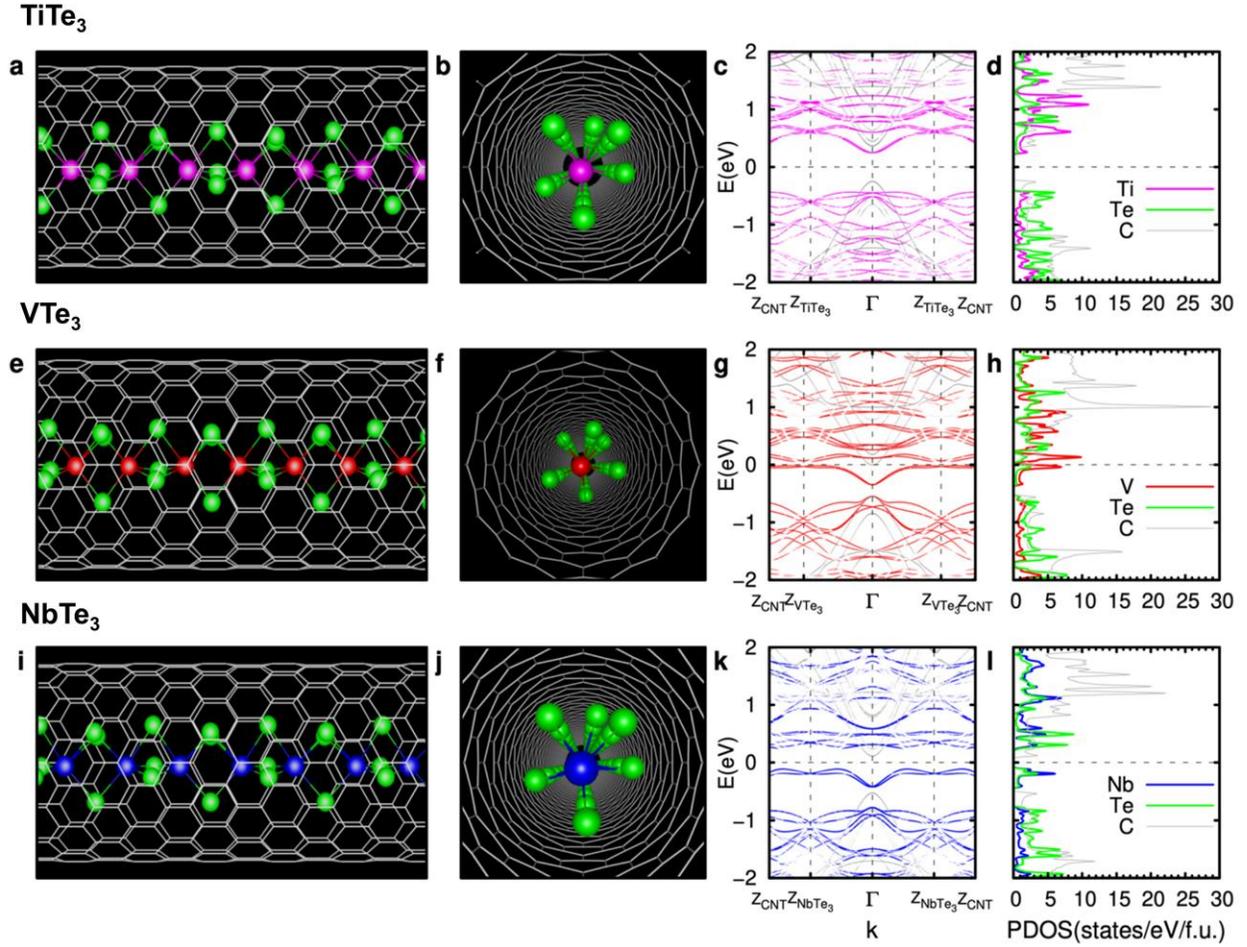

**Figure 4. Calculated atomic structure and electronic structures for TAP single-chain MTe$_3$ (M=Ti, V, and Nb) encapsulated in CNT.** (a-d) TiTe$_3$, (e-h) VTe$_3$, and (i-l) NbTe$_3$. In the 1$^{st}$ and 2$^{nd}$ columns, TAP single-chain encapsulated inside a (14,0) CNT are presented side-on and end-on, respectively. In the atomic structure, the magenta, red, blue, and green spheres represent the Ti, V, Nb, and Te atoms, respectively. In the band structures, the chemical potential is set to zero and marked with a horizontal dashed line. In c, g, and k, the bands represented by magenta, red, blue, and gray lines are projected onto the single-chain TiTe$_3$, VTe$_3$, NbTe$_3$, and CNT, respectively. The bands are then unfolded with respect to the first Brillouin zone of the unit cell of the single chain and the CNT, where zone boundaries for the chain and CNT are denoted as Z$_{MTe3}$ and Z$_{CNT}$, respectively. Here, the lengths of the first Brillouin zones of the TAP chains are half of those in the corresponding TP chains because of the doubled real-space unit cell length of the rocking chains, i.e. $Z_{MTe3} = \pi/2\ b_0^{MTe3}$ where $b_0^{MTe3}$ is the distance between the nearest transition metal atoms. In d, h, and l, the density of states projected onto Ti, V, Nb, Te, and C atoms are presented by magenta, red, blue, green, and gray lines, respectively.

**Table 1. Energetic property of MX₃ bulk.** The Gibbs free energies of formation, $\delta G$, of the MX₃, defined as $\delta G = \epsilon_{MX3} - n_M \epsilon_M - n_X \epsilon_X$, where $\epsilon_{MX3}$, $\epsilon_M$, and $\epsilon_X$ are the total energies per atom of the bulk MX₃, the bulk transition metal M (M = Ti, Zr, Hf, V, Nb, and Ta) and the bulk chalcogen X (X = S, Se, and Te), respectively, and $n_M$ and $n_X$ are the mole fractions of the M and X atoms, respectively. The unit of $\delta G$ is eV/atom.

|    | Ti     | Zr     | Hf     | V      | Nb     | Ta     |
|----|--------|--------|--------|--------|--------|--------|
| S  | -1.042 | -1.110 | -1.082 | -0.650 | -0.666 | -0.690 |
| Se | -0.829 | -0.895 | -0.870 | -0.464 | -0.560 | -0.514 |
| Te | -0.506 | -0.653 | -0.522 | -0.156 | -0.312 | -0.163 |

Supplemental Information for

# Stabilization of NbTe$_3$, VTe$_3$, and TiTe$_3$ *via* Nanotube Encapsulation

Scott Stonemeyer[1,2,3,4†], Jeffrey D. Cain[1,3,4†], Sehoon Oh[1,4†], Amin Azizi[1,3], Malik Elasha[1], Markus Thiel[1], Chengyu Song[5], Peter Ercius[5], Marvin L. Cohen[1,4], and Alex Zettl[1,3,4]*

[1]*Department of Physics, University of California at Berkeley, Berkeley, CA 94720, USA*

[2]*Department of Chemistry, University of California at Berkeley, Berkeley, CA 94720, USA*

[3]*Kavli Energy NanoSciences Institute at the University of California at Berkeley, Berkeley, CA 94720, USA*

[4]*Materials Sciences Division, Lawrence Berkeley National Laboratory, Berkeley, CA 94720, USA*

[5]*National Center for Electron Microscopy, The Molecular Foundry, One Cyclotron Road, Berkeley, CA 94720 USA*

*†These authors contributed equally*

\*Correspondence to: azettl@berkeley.edu

*Electron Microscopy Details:*

After synthesis, filled CNT are cast onto TEM grids (Cu grid, 300 mesh, Lacey Carbon) for characterization. Initial imaging for confirmation and analysis of filling characteristics is done on a JEOL 2010 microscope (TEM, 80 kV). Elemental analysis is done on an aberration-corrected FEI Titan3 (60−300) equipped with a SuperX energy dispersive X-ray spectrometer (EDS, 80 kV, of the few-chain TMT limit). Atomic-resolution STEM imaging for identification of the few chains and TAP single chains is completed at the National Center for Electron Microscopy on TEAM 0.5 which is a Titan 80-300 with a ultra-twin pole piece gap, DCOR probe aberration corrector and was operated at 80 kV and semi-convergence angle of 30 mrad. Images were acquired using the ADF-STEM detector with an inner angle of 60 mrad and a beam current of approximately 70 pA.

STEM simulations were done using EJ Kirkland's autostem program with parameters that matched the experiments. In detail for each simulation, we used a 21.48 Å square simulation box, 80 kV accelerating voltage, 30 mrad convergence semi-angle, 256 pixel sampling, 1 Å slice size, and 50 frozen phonon calculations. After completion, the multislice data was convolved with a 1.3 Å source size to match the contrast seen in the experiment images. Noise was added following Poisson counting statistics to match the 70 pA experimental beam current. This allowed us to interpret the positions of atoms based on the approximate Z-contrast in the images and to compare the projection images to the DFT simulated structures.

*Computational Methods:*

We use the generalized gradient approximation,[19] norm-conserving pseudopotentials,[20] and localized pseudo-atomic orbitals for the wavefunction expansion as implemented in the SIESTA code.[21] The spin–orbit interaction is considered using fully relativistic j-dependent pseudopotentials[22] in the l-dependent fully-separable nonlocal form using additional Kleinman-Bylander-type projectors.[23] We use 1×1×128 Monkhorst-Pack *k*-point mesh for single chains, and about 32 mesh per Å$^{-1}$ of the reciprocal vector for bulk materials. Real-space mesh cut-off of 500 Ry is used for all of our calculations. The van der Waals interaction is evaluated using the DFT-D2 correction.[24] Dipole corrections are included to reduce the fictitious interactions between chains generated by the periodic boundary condition in our supercell approach.[25]

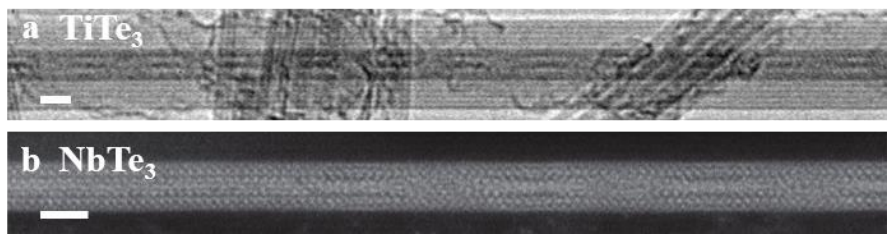

**Figure S1. Tell-tale spiraling of the few-chain limit.** (a) TEM image of few-chain twisting of TiTe$_3$ encapsulated within a MWCNT. (b) STEM image of few-chain twisting of NbTe$_3$ encapsulated within a MWCNT. Scale bars measure 2 nm.

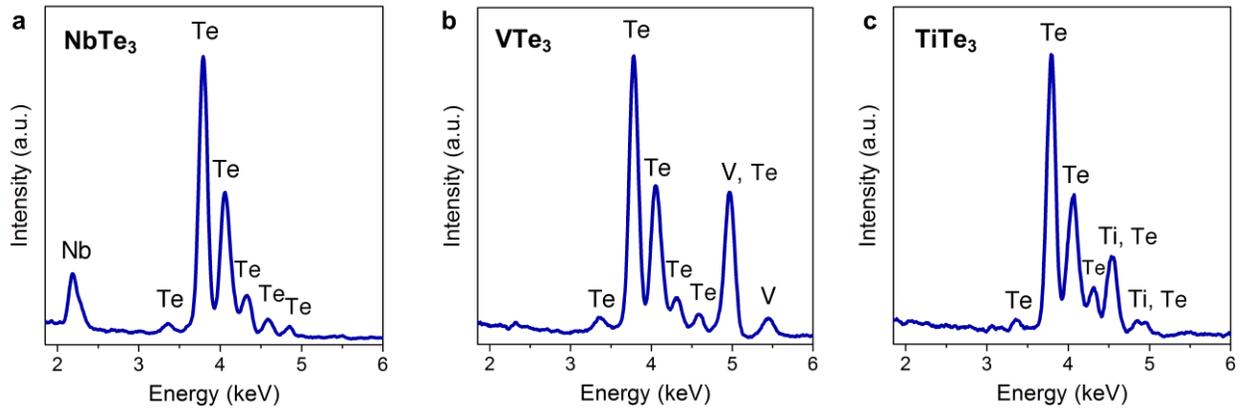

**Figure S2. Energy Dispersive Spectroscopy Spectra.** EDS spectra collected on few-chain specimen for (a) NbTe$_3$, (b) VTe$_3$, and (c) TiTe$_3$, with relevant peaks labeled.

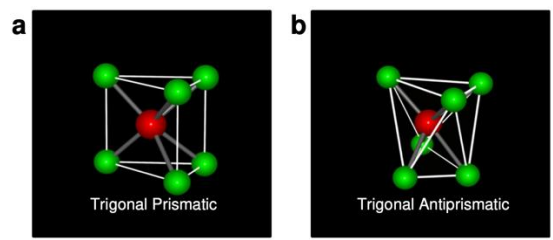

**Figure S3. The geometry of the TP and TAP unit cell.** The basic units of (a) the TP and (b) TAP geometry are shown for comparison.

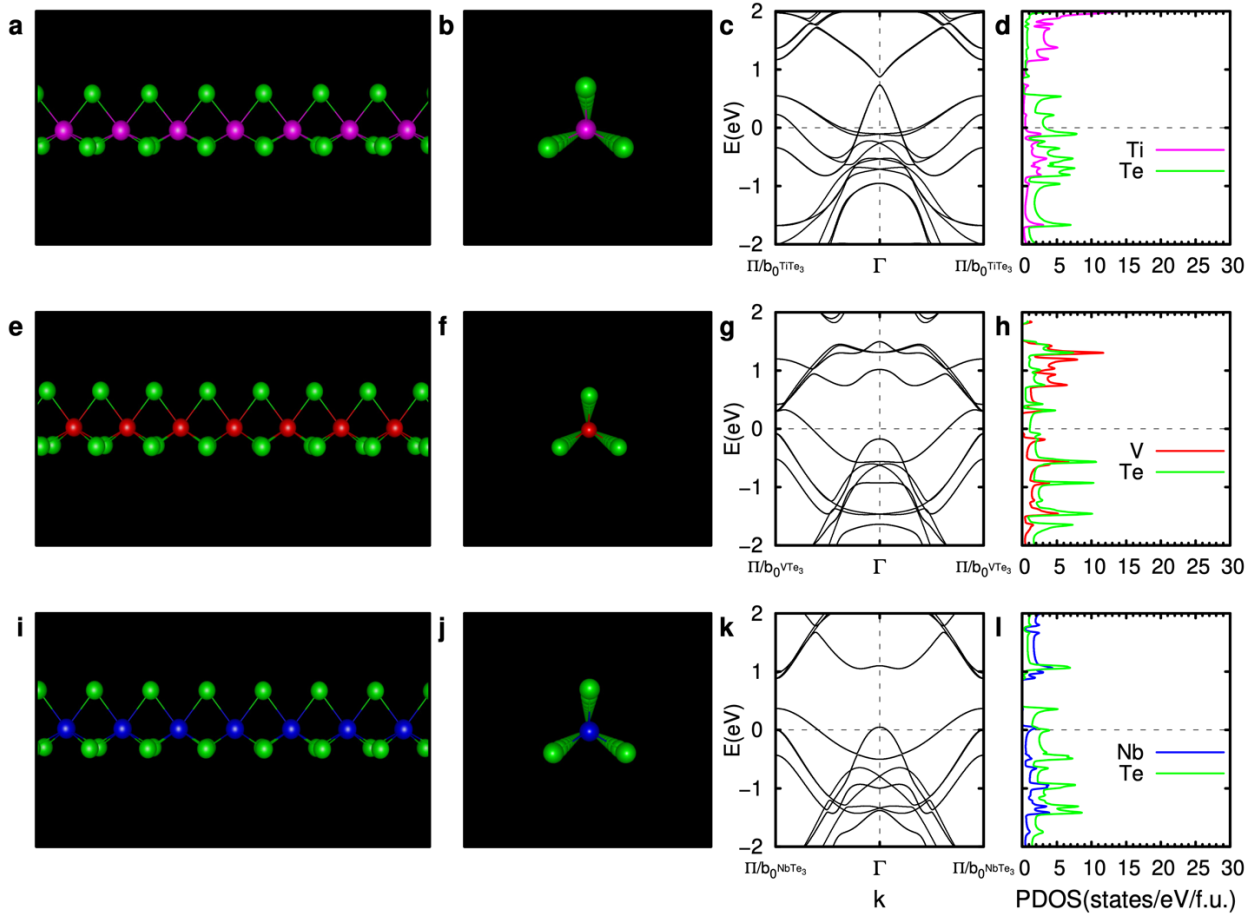

**Figure S4. Calculated atomic structure and electronic structures for TP single-chain MTe$_3$ (M=Ti, V, and Nb) isolated in vacuum.** (a-d) TiTe$_3$, (e-h) VTe$_3$, and (i-l) NbTe$_3$. In the 1$^{st}$ and 2$^{nd}$ columns, TP single-chain isolated in vacuum are presented side-on and end-on, respectively. In the atomic structure, the magenta, red, blue, and green spheres represent the Ti, V, Nb, and Te atoms, respectively. In the band structures, the chemical potential is set to zero and marked with a horizontal dashed line, and the zone boundaries for the chains are denoted as $\pi/b_0^{MTe3}$, where $b_0^{MTe3}$ is the distance between the nearest transition metal atoms. In d, h, and l, the density of states projected onto Ti, V, Nb, and Te atoms are presented by magenta, red, blue and green lines, respectively.

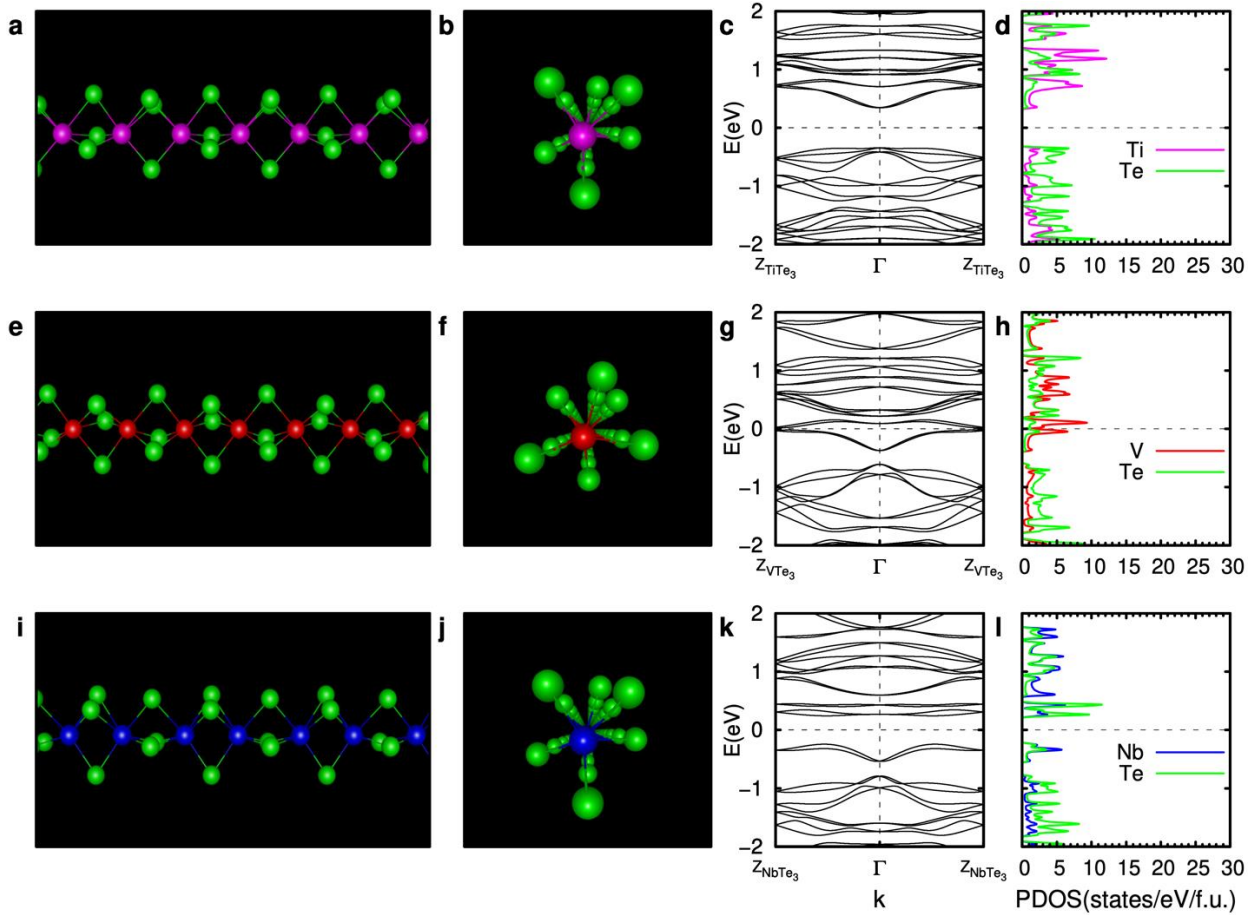

**Figure S5. Calculated atomic structure and electronic structures for TAP single-chain MTe₃ (M=Ti, V, and Nb) isolated in vacuum.** (a-d) TiTe₃, (e-h) VTe₃, and (i-l) NbTe₃. In the 1st and 2nd columns, TAP single-chain encapsulated isolated in vacuum are presented side-on and end-on, respectively. In the atomic structure, the magenta, red, blue, and green spheres represent the Ti, V, Nb, and Te atoms, respectively. In the band structures, the chemical potential is set to zero and marked with a horizontal dashed line, and the zone boundaries for the chains are denoted as $Z_{MTe3}$. Here, the lengths of the first Brillouin zones of the TAP chains are half of those in the corresponding TP chains because of the doubled real-space unit cell length of the rocking chains, i.e. $Z_{MTe3} = \pi/2\, b_0^{MTe3}$ where $b_0^{MTe3}$ is the distance between the nearest transition

metal atoms. In d, h, and l, the density of states projected onto Ti, V, Nb, and Te atoms are presented by magenta, red, blue and green lines, respectively.

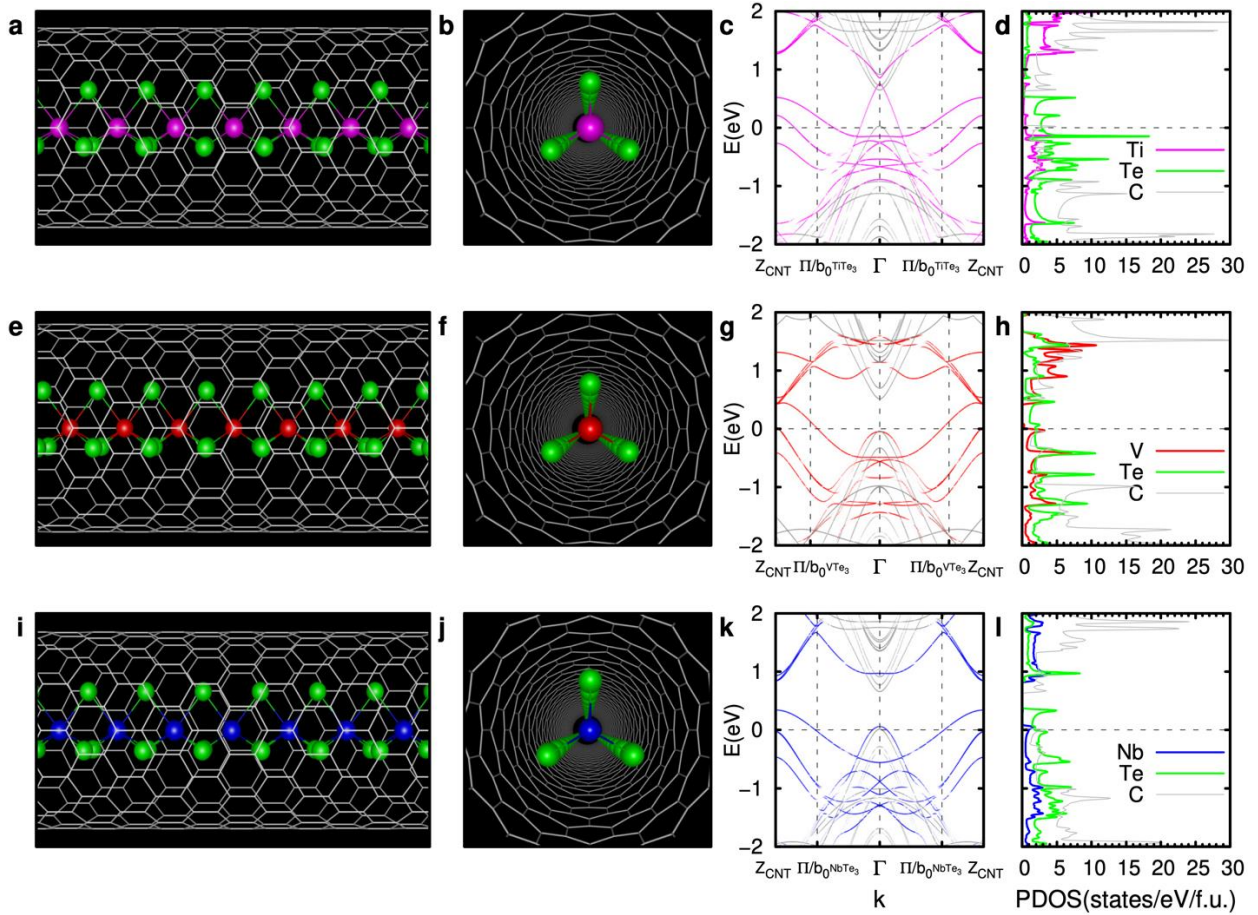

**Figure S6. Calculated atomic structure and electronic structures for TP single-chain MTe$_3$ (M=Ti, V, and Nb) encapsulated in CNT.** (a-d) TiTe$_3$, (e-h) VTe$_3$, and (i-l) NbTe$_3$. In the 1$^{st}$ and 2$^{nd}$ columns, TP single-chain encapsulated inside a (14,0) CNT are presented side-on and end-on, respectively. In the atomic structure, the magenta, red, blue, and green spheres represent the Ti, V, Nb, and Te atoms, respectively. In the band structures, the chemical potential is set to zero and marked with a horizontal dashed line. In c, g, and k, the bands represented by magenta, red, blue and gray lines are projected onto the single-chain TiTe$_3$, VTe$_3$, NbTe$_3$ and CNT, respectively. The bands are then unfolded with respect to the first Brillouin zone of the unit cell of the single chain and the CNT, where zone boundaries for the chain and CNT are denoted as $\pi/b_0^{MTe_3}$ and $Z_{CNT}$, respectively, and $b_0^{MTe_3}$ is the distance between the nearest transition metal

atoms. In d, h, and l, the density of states projected onto Ti, V, Nb, Te, and C atoms are presented by magenta, red, blue, green, and gray lines, respectively.